\documentclass[12pt,a4paper,twoside]{article}

\usepackage{graphicx}
\usepackage{epic}
\usepackage{epstopdf}
\usepackage{amssymb}
\usepackage{amsmath}
\usepackage{sectsty}

\usepackage[utf8x]{inputenc} 
\usepackage[english]{babel}

\sectionfont{\huge}

\begin{document}

\begin{center}
{\huge\bf Modeling the Dynamics of Discussions in Social Networks}
\end{center}

\vspace{0.1cm}
\begin{center}
{\it Petr Kl\'{a}n, Dept. of System Analysis, University of Economics in Prague, Czech Republic, petr.klan@vse.cz}
\end{center}

\vspace{0.7cm}
\noindent
{\bf Abstract:} In this paper, a method of modeling the dynamics of electronic discussions is proposed based on the so called FOPDT model (First Order Plus Dead Time) known from the process control. Knowledge of the model points to possibility of estimating dynamic movements in discussions as well as understanding and designing their maintaining and guidance. Real discussions are processed.

\vspace{1cm}

\section{Introduction}

\begin{figure}[htbp]
\begin{center}
\includegraphics[scale=0.7]{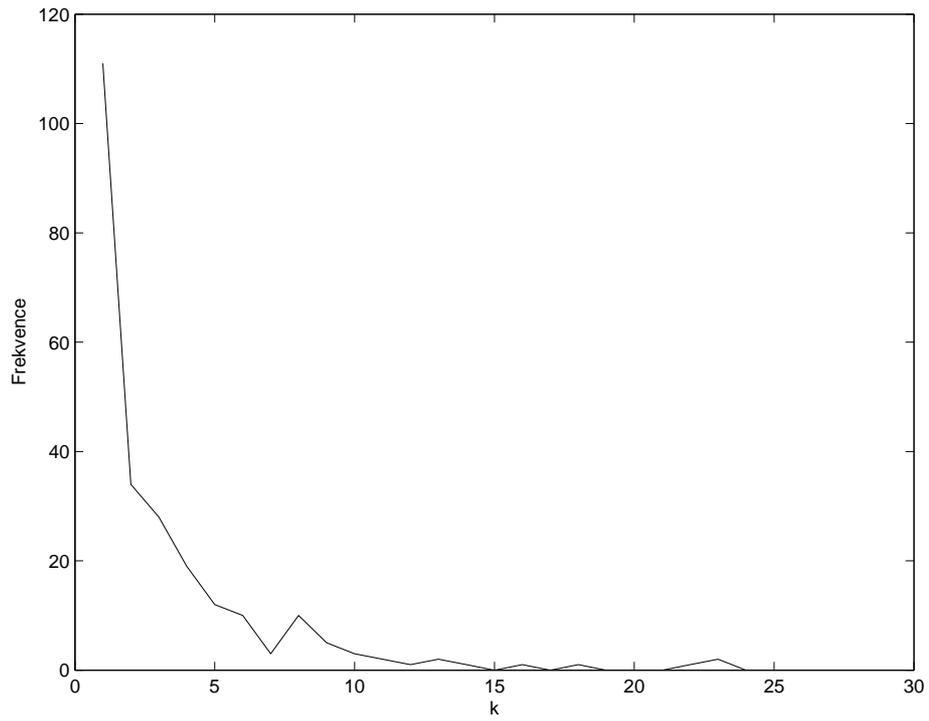}
\end{center}
\caption{\small Illustration of Zipf law. Horizontal axis indicates number of discussion posts and vertical axis frequency of these discussions.} 
\label{zipfq}
\end{figure}

\vspace{0.3cm}\noindent Social networks are associated with complex systems. In order to investigate properties of such systems, statistical models are widely used. They ignore individual network nodes and prefer average behavior of large number of nodes that carry out communication. Furthermore, average characteristics of communication networks are given by Zipf law \cite{pie}. The latter determines, for example, that frequency of Web sites linked from $k$ other sites (in--links) is proportional to the ratio $1 / k^2 $ \cite{mit}. In other words, pages linked from a huge amount of the other Web pages, are very rare. A similar type of dependence can be measured e.g. in the electronic discussion group Chminf--l \cite{chminf}. Frequency of discussions including $k$ discussion posts is approximately proportional to $1/k$. As shown in Fig. \ref{zipfq}, corresponding relationship was measured in the discussion group Chminf--l for year 2010 like in other years. Discussions including more posts than e.g. $15$ are rare in this discussion group. 

\begin{figure}[htbp]
\begin{center}
\includegraphics[scale=0.7]{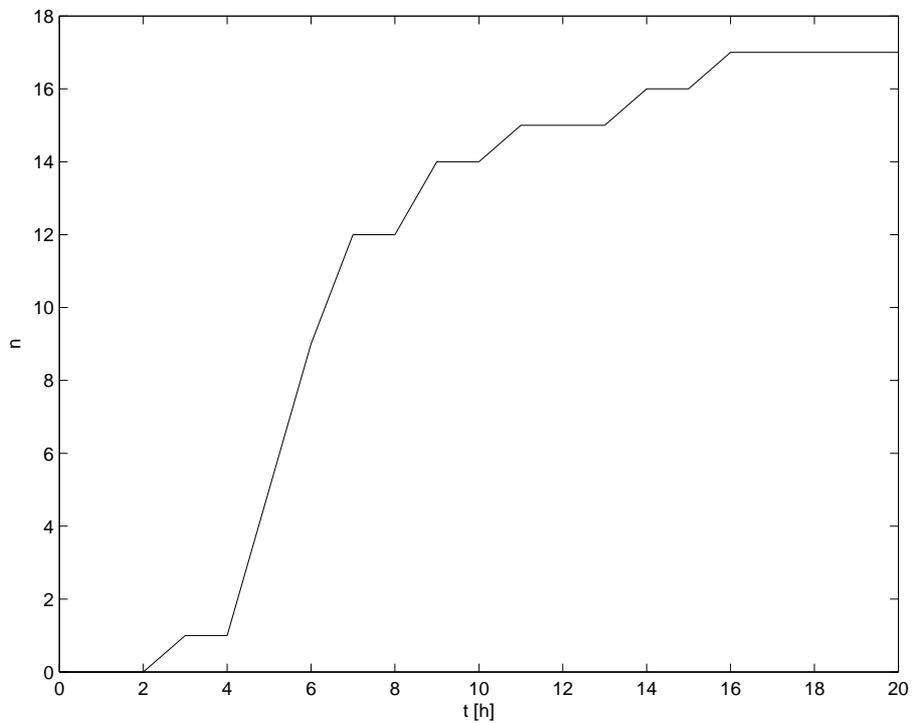}
\end{center}
\caption{\small On the S shape of the logistic curve. Horizontal axis represents time, vertical axis shows growth of discussion post population.}
\label{logcurve}
\end{figure}

\vspace{0.3cm}\noindent In many cases, the patterns of dynamic behavior in complex systems are given by the size changes in some populations $n$, $n>0$, described by differential equations of type $\dot{n}(t)=bn(t)(1-n(t))$. The latter results in well known logistic curves ($ t\geq 0$) 

$$
n(t)=n(0)\dfrac{e^{bt}}{1+n(0)(e^{bt}-1)},
$$
where $b$ denotes size change coefficient and $n(0)$ initial condition \cite{ka}. Logistic curves with the characteristic shapes of the letter S represents a common pattern appropriate to describe the dynamics of complex systems, regardless of their origin. In some cases, cummulative dynamics of responses in electronic discussions exhibits the characteristic shape of the logistic curve too, as illustrated in Fig. \ref{logcurve} for Chminf--l discussion carried out on July 22, 2011. 

\begin{figure}[htbp]
\begin{center}
\includegraphics[scale=0.7]{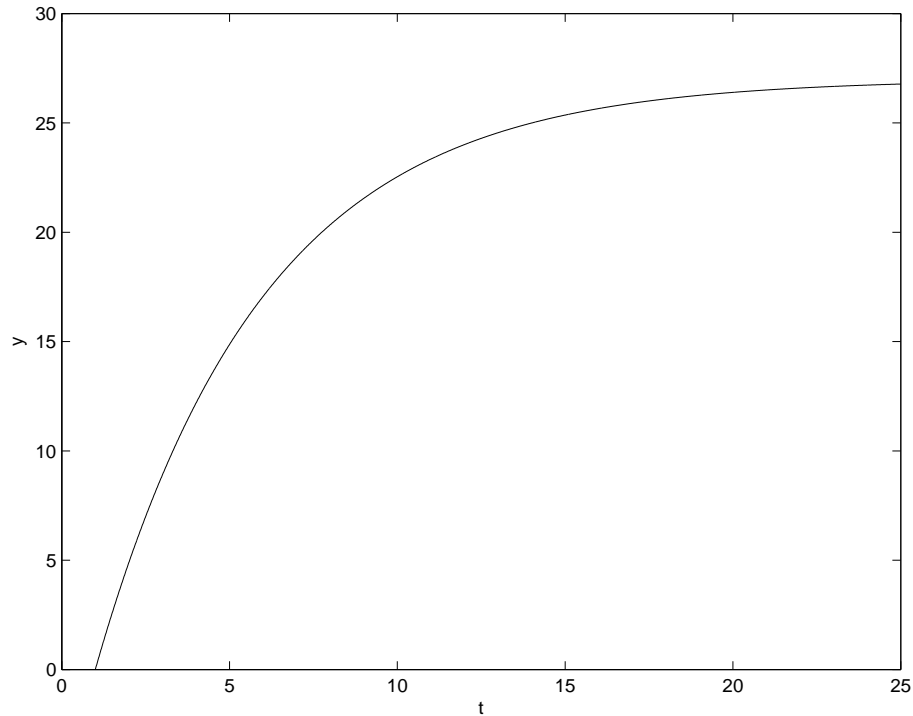}
\end{center}
\caption{\small Three--parameter model step response.}
\label{exp}
\end{figure}

\vspace{0.3cm}\noindent Advanced methods for PID (Proportional $+$ Integral $+$ Derivative) control use mathematical process models which have a potential to study dynamic events related to communication in social networks as well. Primarily, there are three--parameter models \cite{kg1} or FOPDT models (First Order Plus Dead Time) with the corresponding differential equation  $T\dot{y}(t)+y(t)=Ku(t-L)$ represented by the transient solution ($t\geq L$)

$$
y(t)=K\left(1-e^{\dfrac{-(t-L)}{T}}\right)
$$
given by the unit step response for $y(0)=0$. An example of such transient response with $K=27$, $T=5$ and $L=1$ is shown in Fig. \ref{exp}. Variables $y,u$ indicate input and output and non--negative parameters $K,T,L$ gain, time constant and time delay, respectively. Physical units are not specified in this illustrative example. Using Laplace transform, this model corresponds to a transfer function

$$
K\dfrac{e^{-sL}}{Ts+1}.
$$

\vspace{0.3cm}\noindent The paper shows that the dynamics of communication networks, under certain conditions, can be identified by the three--parameter models. The latter is yet suitable for purposes of prediction and management of discussions using principles associated with PID control. Since the three--parameter models are widely used in automatic control, various rules have been developed for their use. Here, knowledge of parameters $K, L, T$ is decisive. Just widely known model, the intuitive clarity of its meaning, identification and use allow to better understand the communication dynamics in social networks, including such elements of communication as posting, liking etc. For example,  use of the specific value of $63 \%$ of $K$ at $t=T+L$ is available for estimating sizes of discussions.\\
\\ 
The following section explains motivation for use of three--parameter model for discussion dynamics description. Further sections are concerned with identification of three--parameter models from cummulative data of discussions. Moreover, some real examples of identification in electronic discussion group Chminf--l are introduced.

\section{Motivation}

\begin{figure}[htbp]
\begin{center}
\includegraphics[scale=0.35]{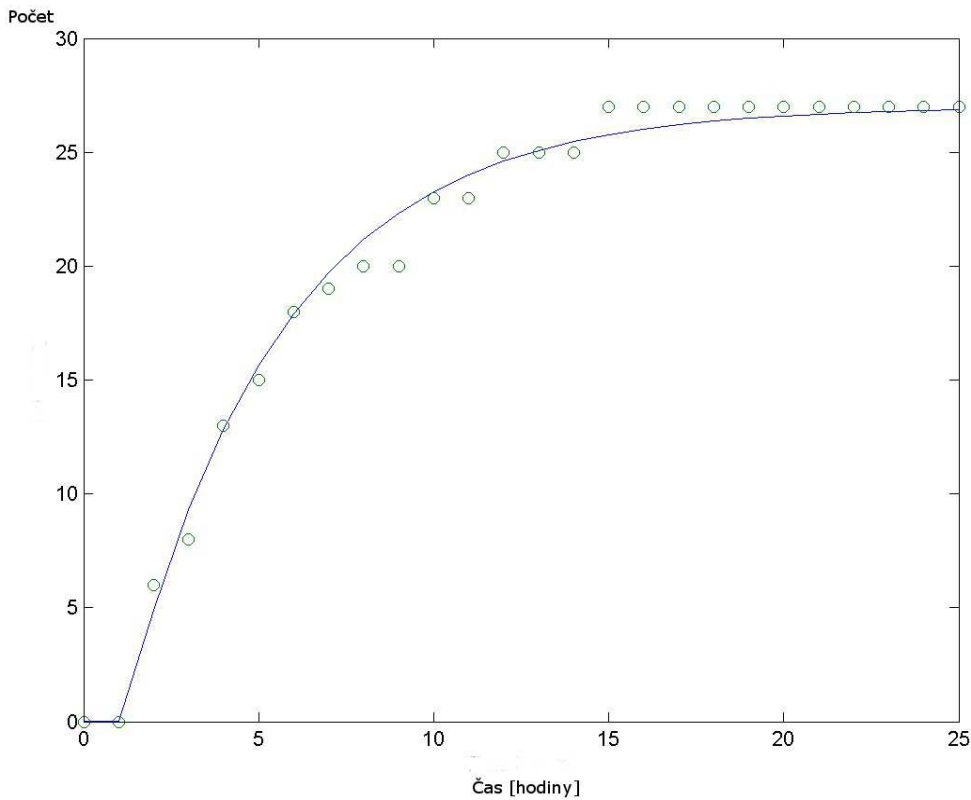}
\end{center}
\caption{\small Dynamics of discussion on Wikipedia decline.}
\label{decline}
\end{figure}

\vspace{0.3cm}\noindent The author was a participant in one of the discussions on social network, which debated the topic of possible Wikipedia decline despite the noble goal of Wikipedia to assemble all human knowledge. Although a unique system in the history of civilization, the number of active Wikipedia contributors constantly decreases significantly. The principal reason is considered in constraints that community of Wikipedia contributors has created to discourage new contributors. \\
\\
Discussion had a classical format. At the beginning it was an initial post about Wikipedia decline giving the link to the paper \cite{sim} to which the participants debated. After the end discussion, the author collected the dynamics of discussion, i.e. the cumulative number of posts $y$ to vertical axis versus related times $t$. It results in Fig. \ref{decline} where the circle indicates the cumulative number of discussion posts at the time. The first post on the subject appeared after about an hour from the initial contribution (transport delay). It is followed by the relatively steep growth of posts to gradually slowed after $24$ hours to reach the final steady state of $27$ posts.\\
\\

\vspace{0.3cm}\noindent If observing dynamics of this discussion then one can not ignore the similarity with the exponential dependence of transient response in Fig. \ref {exp}. The same transient response with parameters $K=27$ posts/time, $T=5$ hours and $L=1$ hour is inscribed by continuous line in Fig. \ref{decline}. It well characterizes and describe the dynamics of discussion and associates with a principal question if is it possible to describe discussions by the related transfer function  

$$
27\dfrac{e^{-s}}{5s+1}
$$
or whether the three--parameter models describe the dynamics of electronic discussions in social networks and the Internet in general?

\section{Discussion dynamics identification}

\begin{figure}[htbp]
\begin{center}
\includegraphics[scale=0.7]{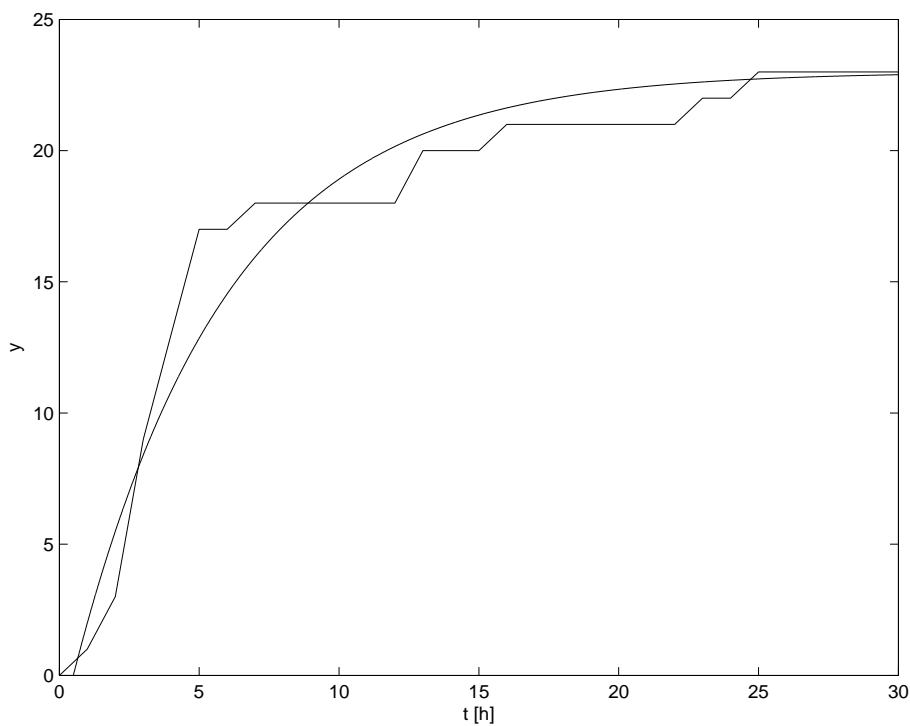}
\end{center}
\caption{\small Dynamics of discussion in Chminf--l and its identification by three--parameter model.}
\label{chem1}
\end{figure}

\vspace{0.3cm}\noindent At the beginning, it is necessary to answer the question whether the dynamics of discussion behaves as a transient response to step change? Yes, it does. The initial post from which the debate unfolds, appears and lasts. Thus, it fulfills the same 'shock' role like unit step expressed by  $1/s$ in the Laplace transform. At the time of this step or the initial post, it is possible to set up the start time $t=0$ and (with zero initial condition) to start measurement of transient response as the cumulative number of the posts. 
\\
\\
The dozens of discussions were examined to verify if the discussion dynamics captured in Fig. \ref{decline} does not represent a random but repeatable shape. For this purpose, the freely accessible archive of electronic discussion group for chemical informatics Chminf--l (https: // list.indiana.edu) was selected. The author is a member of this group since 1995 \cite{kmsr}. The group has about 1500 participants and well serve as a model paradigm of communication in social networks because indicating the same patterns introduced by the Zipf law.
\\
\\
Discussion subjected as `What are faculty instead of using chemistry reference titles?' from February 2010 represents a characteristic example. The initial post (unit step) was sent at 16:35 ($t=0$). The total number of the following posts was $23$, which immediately represents gain $K$ associated with the unit step. The dynamics of the discussion is captured in Fig. \ref{chem1} (irregular curve) together with the unit step response of identified three--parameter model by $K= 23$ posts/time, $T=5.5$ hours and $L=0.5$ hour. In order to obtain these parameters, the algorithm of identification described in \cite{kg1} was used. Therefore, discussion is identified by the transfer function

$$
23\dfrac{e^{-0.5s}}{5.5s+1}.
$$

\vspace{0.3cm}\noindent Discussions with a large number of posts (typically $> $ 10) indicate common patterns in the aforementioned public archive. It examines discussions in private social networks, too. Data from large amount of discussions confirm the hypothesis that the dynamics of the discussion can be interpreted (with sufficient accuracy) as the dynamics of the three--parameter model. In other words, dynamics of a social network discussion can be sufficiently represented by three parameters, $K, L$ and $T$. 
\\
\\
Moreover, it appears that while the parameters of $L$ and $T$ in the dynamics of the discussions are not too much different (typical delay currently was within one hour and the time constant from $1$ to $5$ hours), the parameter $K$ representing the cummulative number of posts significantly varies. Thus, the situation relates to dynamic systems with a highly variable gain. On the other hand, the Fig. \ref{zipfq} suggests that discussions with gain of more than $K=15$ are significantly rare in the investigated discussions.

\section{Examples of identification}

\begin{figure}[htbp]
\begin{center}
\includegraphics[scale=0.7]{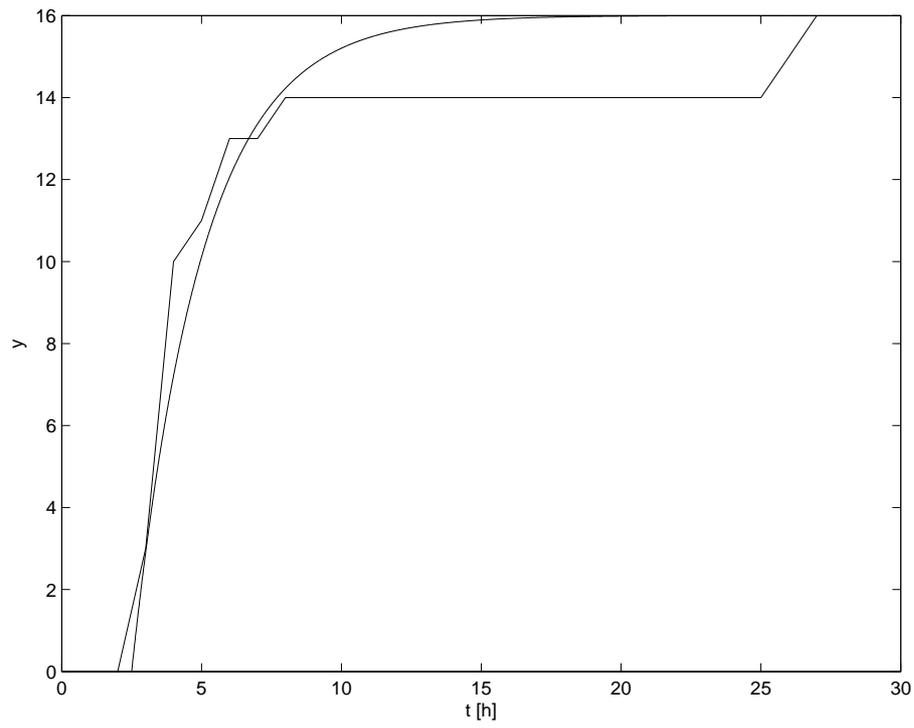}
\end{center}
\caption{$ 16\dfrac{e^{-2.5s}}{2.5s+1}.$} \label{chem2}
\end{figure}

\vspace{0.3cm}\noindent This section shows examples of identification and illustrates situations met in the real practice of discussions. 
\\  
\\
{\bf Example 1}: Chminf--l 10.3. 2011, beginning: time 15:59, subject: ISTL article \ldots, basic time unit: hour, Fig. \ref{chem2}.\\
\\
It is a case of the discussion, which had the standard dynamics and that the last two posts were added after many hours. In the case of international discussions due to different time zones (or e.g. working absence), it is a common phenomenon and it currently represents $10\%$ error given by a noise. It is known \cite{kg2} that, for example, good PID tuning has to deal with $20\%$ changes of the controlled process parameters.

\begin{figure}[htbp]
\begin{center}
\includegraphics[scale=0.7]{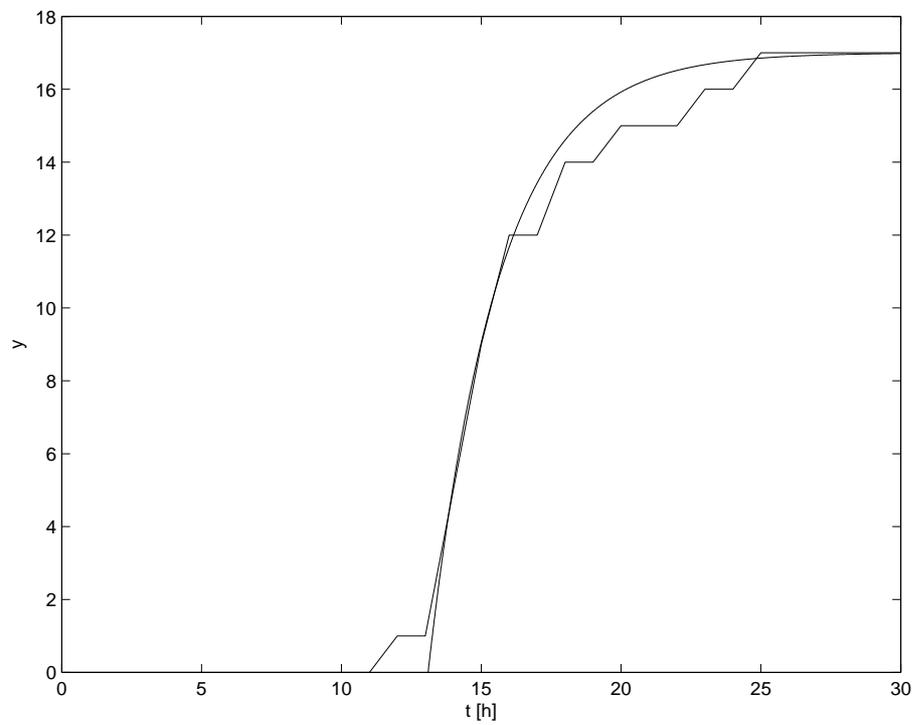}
\end{center}
\caption{$ 17\dfrac{e^{-13.1s}}{2.5s+1}.$} \label{chem3}
\end{figure}

\vspace{0.3cm}\noindent {\bf Example 2}: Chminf--l 6.1. 2012, beginning: time 1:41, subject: Authors listed in search of ACS Pubs, basic time unit: hour, Fig. \ref{chem3}.\\
\\
The example shows the situation with an extremely large time delay, the process by which the first contribution post to the discussion came after about 12 hours. So long delay is examined in discussions rare and when they do, it is usually given by inadequate time posting of the first discussion post.

\begin{figure}[htbp]
\begin{center}
\includegraphics[scale=0.7]{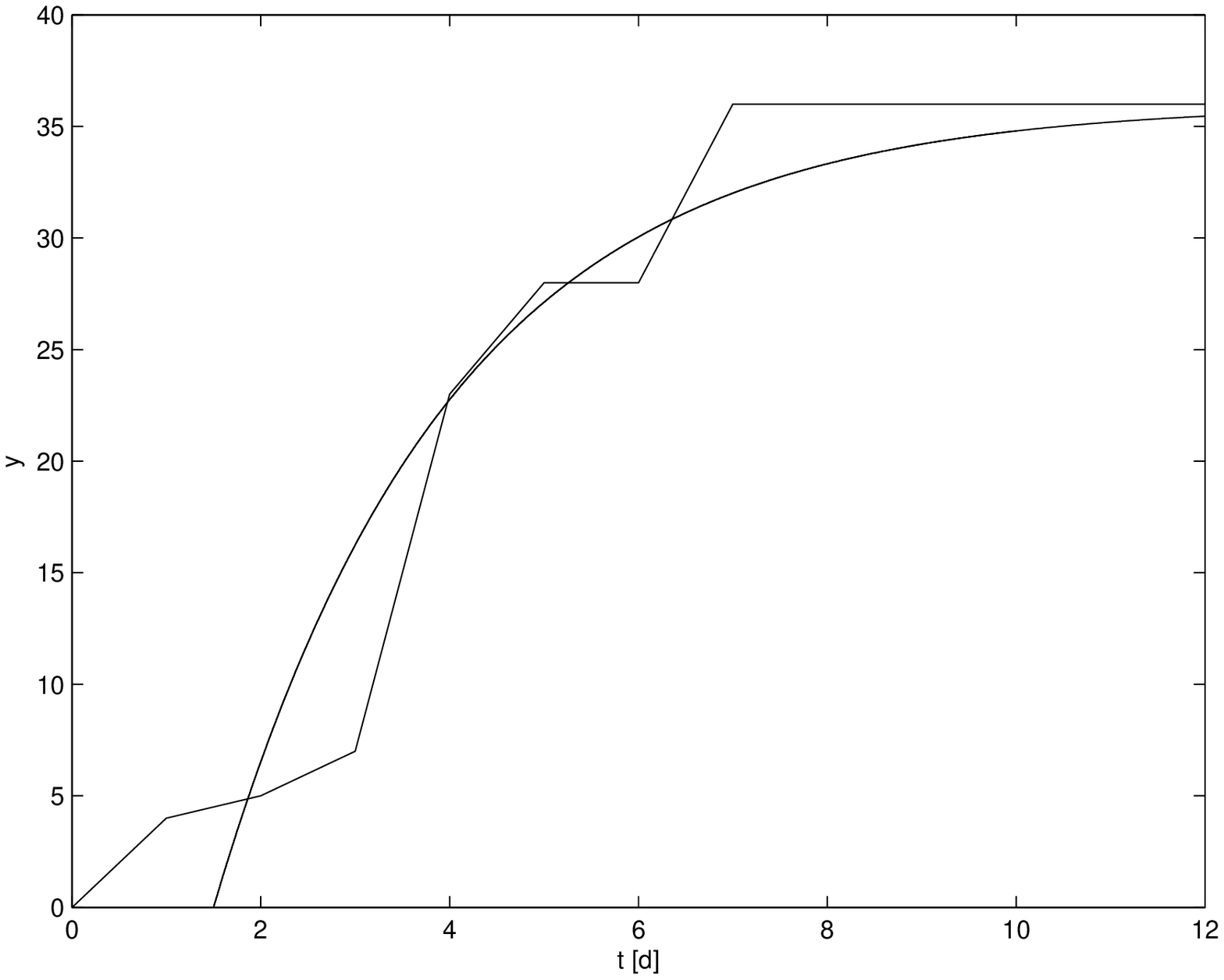}
\end{center}
\caption{$ 36\dfrac{e^{-1.5s}}{2.5s+1}.$} \label{chemd}
\end{figure}

\begin{figure}[htbp]
\begin{center}
\includegraphics[scale=0.7]{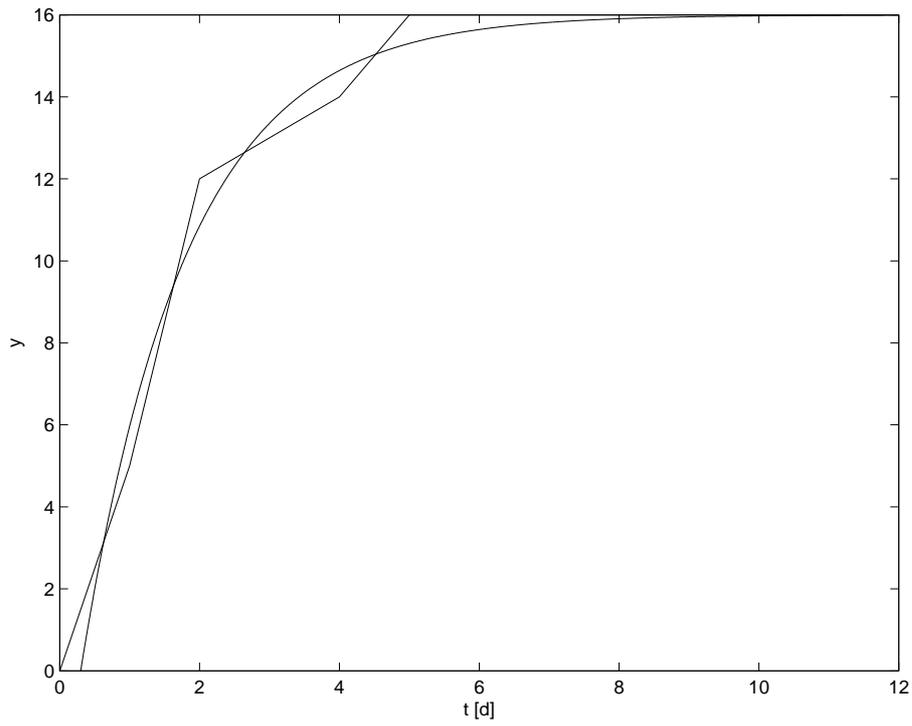}
\end{center}
\caption{$ 16\dfrac{e^{-0.3s}}{1.5s+1}.$} \label{chemd1}
\end{figure}

\vspace{0.3cm}\noindent {\bf Example 3}: Chminf--l 22.7. 2011, beginning: time 17:23, subject: SciFinder Piracy, basic time unit: day, Fig. \ref{chemd}. Chminf--l 7.11. 2011, beginning: time 20:03, subject: Handbook 92d edition \ldots some thoughts, basic time unit: day, Fig. \ref{chemd1}.\\
\\
It is illustration of discussions with long time delays and time constants (in days). Furthermore, Fig. \ref{chemd} captures a somewhat chaotic shape of the discussion dynamics, in which, despite the large variations, it is possible to recognize the exponential character (and finally the logistic curve).

\begin{figure}[htbp]
\begin{center}
\includegraphics[scale=0.7]{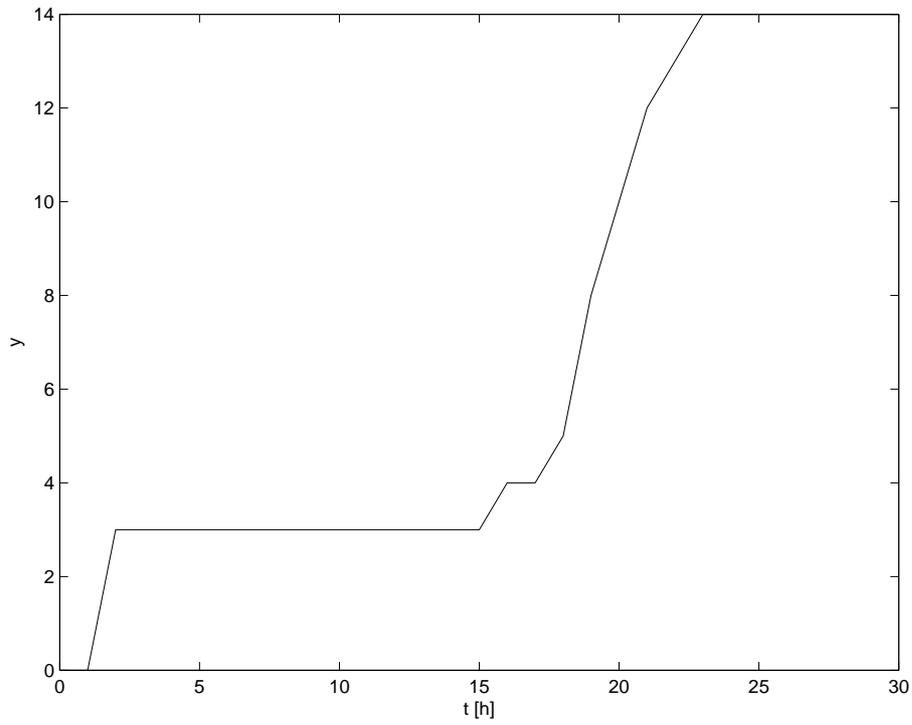}
\end{center}
\caption{\small Discussion under work absences.} \label{chem_noc}
\end{figure}

\vspace{0.3cm}\noindent {\bf Example 4}: Chminf--l 22.6. 2011, beginning: time 18:13, subject: Call for papers from Journal of Chemistry and Chemical Engineering, basic time unit: hour, Fig. \ref{chem_noc}.\\ 
\\
Capturing a break dynamics discussion (e.g. due to the absence in work), which may occur in the discussion, unlike the technological process. It is interesting to see that the dynamics of the discussion is interrupted while the delay continues. 

\begin{figure}[htbp]
\begin{center}
\includegraphics[scale=0.7]{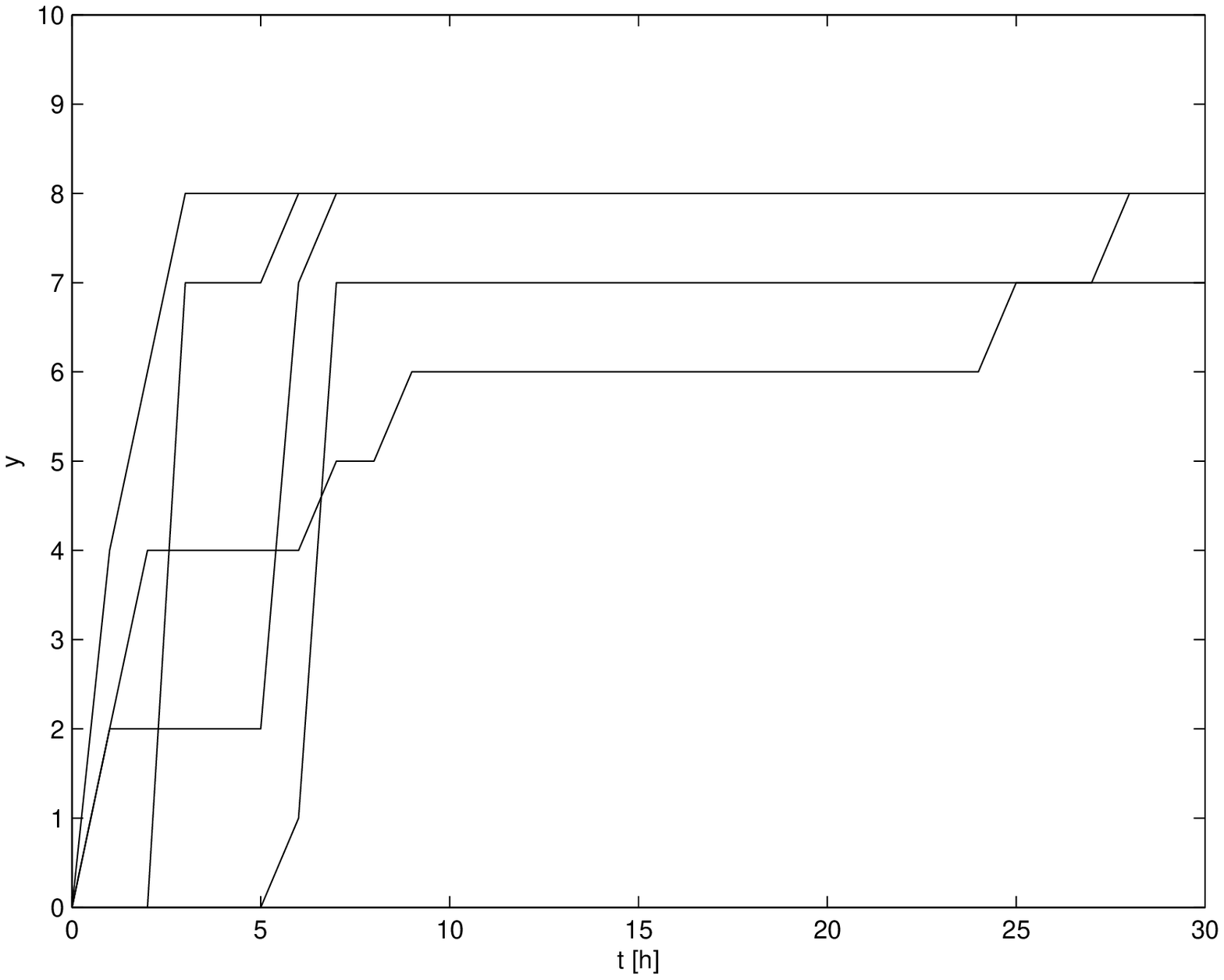}
\end{center}
\caption{\small Discussions with a small number of posts.} \label{k8}
\end{figure}

\vspace{0.3cm}\noindent {\bf Example 5}: Chminf--l 2010, first 5 discussions including 8 posts, Fig. \ref{k8}.\\
\\
Fig. \ref{k8} shows examples of the discussion dynamics with a small number of posts. Such discussions are frequently included in the Chminf--l. In one case, the last post came even after more than $30$ hours from the initial post. In Fig. \ref{k8}, a step response without time delay is experienced as well as a delayed response with extremely small time constants as well as response with a time constant of around $10$ hours. All of these responses can be identified by three--parameter model similarly as in previous cases. However, using the predictive power of the model, i.e. prediction of the characteristic time $L+T$ is difficult. On the other hand, the gain $K$ for newly starting discussions can be estimated by the Zipf law (well introduced eg. in \cite{pie}). With each new discussion will be possible, based on the subject of discussion, to estimate its probable order associated with the number of posts.

\section{Conclusion}

\vspace{0.3cm}\noindent The paper shows an innovative use of three--parameter FOPDT models in identifying dynamics of social discussions. By this view, the cummulative dynamics of discussions is described as e.g. dynamics of mechanical systems. The paper gives examples of identification of real discussions in the electronic discussion group called Chminf--l. \\
\\
It is shown that dynamics of discussions is well described by the three parameter FOPDT model known from the process control. At the same time, the predictive ability of three--parameter models can be used. In the case of a small number of posts, it is possible to use a combination of the model with the effects of Zipf law. \\
\\
In addition to the prediction and estimation of dynamic properties of discussions, the three--parameter model serves in managed discussions when calculating proportional and integral parts. The proportional part can manifest itself by inserting proportionally opposing views or to removing the offending posts, the integral part then historical experience on the subject, experience from previous discussions, etc.

\end{document}